\documentclass[a4paper,preprintnumbers,floatfix,superscriptaddress,prl,twocolumn,showpacs,notitlepage,longbibliography]{revtex4-1}
\usepackage[utf8]{inputenc}
\usepackage[T1]{fontenc}
\usepackage[sc,osf]{mathpazo}
\usepackage{amsmath, amsthm, amssymb,amsfonts,mathbbol,amstext}
\usepackage{graphicx}
\usepackage{dcolumn}
\usepackage{bm}
\usepackage{bbm}
\usepackage{hyperref}
\usepackage{mathtools}
\usepackage{comment}
\usepackage{color}
\usepackage{multirow}
\usepackage{diagbox}
\usepackage{float}

\def\1{\mathbf{1}}
\def\0{\mathbf{0}}
\def\o{\mathbb{O}}

\newcommand{\ie}{{\it{i.e.~}}}

\newcommand{\processnext}[1]{%
  \ifx\listfinish#1\empty\else\listact{#1}\expandafter\processnext\fi}

\begin{document}

\title{Statistical properties of the quantum internet}

\author{Samura\'i Brito}
\affiliation{International Institute of Physics, Federal University of Rio Grande do Norte, 59070-405 Natal, Brazil}
\author{Askery Canabarro}
\affiliation{International Institute of Physics, Federal University of Rio Grande do Norte, 59070-405 Natal, Brazil}
\affiliation{Grupo de F\'isica da Mat\'eria Condensada, N\'ucleo de Ci\^encias Exatas - NCEx, Campus Arapiraca, Universidade Federal de Alagoas, 57309-005 Arapiraca-AL, Brazil
}
\author{Rafael Chaves}
\email{rchaves@iip.ufrn.br}
\affiliation{International Institute of Physics, Federal University of Rio Grande do Norte, 59070-405 Natal, Brazil}
\affiliation{School of Science and Technology, Federal University of Rio Grande do Norte, 59078-970 Natal, Brazil}
\author{Daniel Cavalcanti}
\email{daniel.cavalcanti@icfo.eu}
\affiliation{ICFO-Institut de Ciencies Fotoniques, The Barcelona Institute of
Science and Technology, 08860 Castelldefels (Barcelona), Spain}

\begin{abstract}
Steady technological advances are paving the way for the implementation of the quantum internet, a network of locations interconnected by quantum channels. Here we propose a model to simulate a quantum internet based on optical fibers and employ network-theory techniques to characterize the statistical properties of the photonic networks it generates. Our model predicts a continuous phase transition between a disconnected and a highly-connected phase characterized by the computation of critical exponents. Moreover we show that, although the networks do not present the small world property, the average distance between nodes is typically small.
\end{abstract}

\maketitle
Network science is a multidisciplinary field that offers a common language to study statistical properties of a variety of systems such as social, biological, and economical networks \cite{barabasi2016network}. On the basis of its success is the fact that systems are seeing simply as graphs, i.e. a set of nodes interacting via edges. In this approach it is not the particular working or behavior of the individual constituents that matters, but how connected they are. This viewpoint led to the discovery that systems that are very different in nature, such as the internet, scientific collaborations, or protein networks, are very similar from a network perspective. Furthermore, understanding the network connectivity allows to design better man-made networks, such as power grids, transport networks or company organisation. 

A new type of communication network, the quantum internet, is currently under development \cite{kimble2008quantum,wehner2018quantum}. It consists of distant parties connected by quantum channels through which quantum bits can be exchanged. This new network will boost our capabilities of communication by allowing the execution of protocols which are more efficient than their classical counterpart, or that have no classical analog whatsoever. The main example of such advantage is the possibility of securing messages with quantum cryptography \cite{Gisin_2002}, currently one of the most advanced quantum technologies. Other anticipated applications are clock synchronisation \cite{ClockSinc} and private quantum computation on a cloud \cite{Broadbent_2009,PhysRevA.96.012303}. From a fundamental perspective, quantum networks will also allow us to reach physical phenomena that have no classical analog. An example is the distribution of entanglement across the network, which will allow distant parties to perform quantum teleportation or to establish correlations with no classical explanation and defy our notions of causality \cite{Brunner_2014}.

Here we propose a model to simulate the quantum internet assuming that it is going to be built from optical fibers. We use this model to predict global properties of typical photonic networks, such as their connectivity, nodes distance, and aggregation. Our findings predict a phase transition in the network connectivity as a function of the density of nodes: there is a critical density above which the network changes from being disconnected to presenting a giant connected cluster. We estimate the value of this critical density and the critical exponents characterising the phase transition. Nicely, few nodes are needed to make photonic networks of realistic sizes fully connected. However, as opposed to the current internet \cite{Albert1999}, the quantum internet does not present the small world property. Notwithstanding, the typical network distances between nodes are small, implying that few entanglement swappings have to be employed to distribute entanglement between any two nodes.

\begin{figure}[t!]
\begin{center}
\includegraphics[width=1\columnwidth]{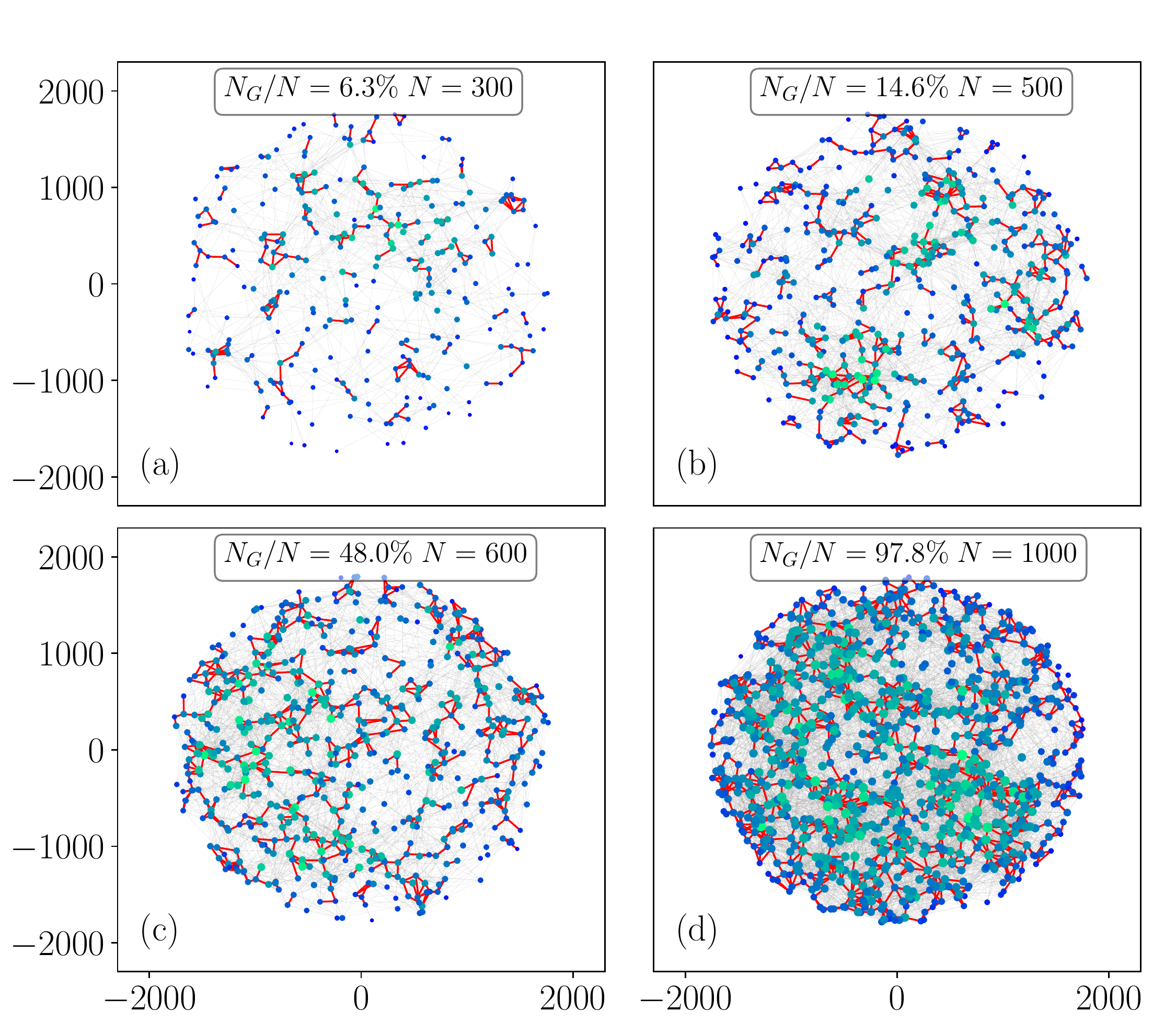}
\end{center}
\caption{\textbf{Samples from the quantum internet}. The grey edges represents the fiber-optics networks generated at step 1 (see main text). The red edges show the photonic links established in step 2. Greener (bluer) nodes are more (less) connected, following a Poisson distribution. $N_G$ refers to the number of nodes belonging to the biggest cluster in the network, and $N$ the total number of nodes. The plots considered $R=1800$ km (giving, approximately, the US area) and show that the biggest cluster consists of $97,8\%$ of the nodes when $N=1000$.}
\label{network02}
\end{figure}

Formally, a network model is defined by a set of $N$ nodes being connected by vertices according to a given probabilistic rule. The central goal of network science is to understand the asymptotic properties of networks as the number of nodes increases. A particularly relevant example is given by random networks \cite{erdos1959random, erdos1960evolution}, defined by a model where every pair of node is connected with probability $p$ \cite{gilbert1959}. The characteristic trait of random networks is that for sufficiently large number of nodes $N$, the probability of finding a node with $k$ connections $P(k)$, called the degree distribution, can be approximated by a Poisson distribution $P(k) = \frac{e^{-\langle k \rangle} \langle k \rangle^k}{k!}$, where $\langle k \rangle=p(N-1)$ is the average connectivity of the network. Despite being very simple, the random network model presents very rich statistical phenomena. For instance, it displays a phase transition: there is a critical probability $p_c$ such that if $p < p_c$ the network is composed by small and disconnected clusters and, if $p > p_c$ a giant cluster with size of same order of the whole network is present. Another striking feature is the appearance of a phenomenon known as \textit{small-world}. This refers to the fact that the average shortest path length (\ie the shortest path between two nodes) scales logarithmically with $N$, meaning that the typical distances between pairs of nodes is very short compared to the size of the network.  

Another important property of networks is the average clustering coefficient. It captures how the neighbors of each node are connected between them on average. Let us first define the local clustering coefficient of node $i$ as $C_i=\frac{2n_i}{k_i(k_i - 1)}$, where $n_i$ the number of edges between the $k_i$ neighbours of the site $i$ and $k_i(k_i - 1)/2$ is total possible number of edges between them. If $C_i=0$ there is no links between the neighbors of $i$, while $C_i=1$ indicates that the neighbors of $i$ form a fully connected graph. The average clustering coefficient is defined as $\langle C \rangle = \frac{1}{N}\sum_{i} C_i$. For random networks $\langle C \rangle =  \langle k \rangle / N$, showing a decrease with the network size. 

In what follows we will propose a model to simulate the quantum internet and use it to predict these properties for photonic networks. As we will see, these networks present similarities and differences with the random networks.

Our model considers a network built from optical fibers, the main candidate to carry quantum information encoded in photons. Other technologies, such as quantum satellites, are also being considered and will probably be combined with the fiber-optics infra-strucure \cite{Yin1140,Liao2018}. Thus, the results presented here can be seen as benchmark to be improved by additional technologies. Our model is defined by the following steps:

\emph{Step 1 - Fiber-optics network simulation.}  We first distribute $N$ nodes uniformly in a disk of radius $R$ (points at Fig. \ref{network02}) \footnote{Different geometries and position distribution could be easily considered.}. Following \cite{lakhina2003}, we simulate how the optical fibers are distributed among these nodes using the Waxman model \cite{waxman1988}, which considers that each pair of nodes $i$ and $j$ are connected by a fiber (grey lines at Fig. \ref{network02}) with probability given by $\Pi_{ij}=\beta e^{-d_{ij}/\alpha L}$, where $d_{ij}$ is the Euclidean distance between $i$ and $j$, $L$ is the maximum distance between any two nodes, the parameter $\alpha>0$ controls the typical edge length of the network (the maximum distance of two nodes directly connected), and $0<\beta \leq 1$ controls the average degree of the network. The constants $\alpha$ and $\beta$ have been estimated for particular optical fiber networks, such as for the US fiber-optics network where $\alpha L=226$ km and $\beta=1$ \cite{lakhina2003, Durairajan2015}. We will use these values in the numerical simulations presented here. 

\emph{Step 2 - Photonic network simulation.} Once we generate the fiber-optics network we simulate the transmission of photons through it. It turns out that photonic losses increase exponentially with the fiber length \cite{Gisin2015}. More precisely, the transmissivity determining the fraction of energy received at the output of a fiber link connecting nodes i and j is given by $p_{i j} = 10^{-\gamma d_{i j}/10}$, where $d_{ij}$ (km) is the Euclidean distance between $i$ and $j$ and the value of the fiber loss $\gamma$ depends on the photon wavelength. For instance, for the silicon fiber, losses are minimized at the wavelength of $1550$nm, achieving $ \gamma \simeq 0.2$ dB/km, the value we consider in our simulations. Even with further advances, the intrinsic physical loss limit of the silica optical fibers is estimated to be between $0.095$ to $0.13$ dB/km ~\cite{TSUJIKAWA2005319}. 

Finally we define the probability $P_{i j}$ that two nodes are connected as
\begin{equation}
    P_{i j} = 1-(1-p_{i j})^{n_p} \label{pij}.
\end{equation}
The free parameter $n_p$ controls how many photons are sent between each node in the attempt of generating a photonic link, \ie two nodes are connected if at least one out of $n_p$ photons is transmitted between them. For a illustrative matter, we chose $n_p=1000$ in the figures depicted here, as this value guarantees that connections over 100km, the order of the  state-of-art quantum communication experiments, are established. We highlight, however, that extensive simulations have been also performed with different values (see Appendix), showing that the qualitative features described below of the photonic networks are universal and independent of the value of $n_p$.

We repeat steps 1 and 2 above $10^3$ times to generate different instances of the quantum photonic internet and calculate its typical properties. Some samples of the networks generated by this algorithm are shown in the Fig.~\ref{network02}.

\begin{figure}[t!]
\begin{center}
\includegraphics[scale=.41]{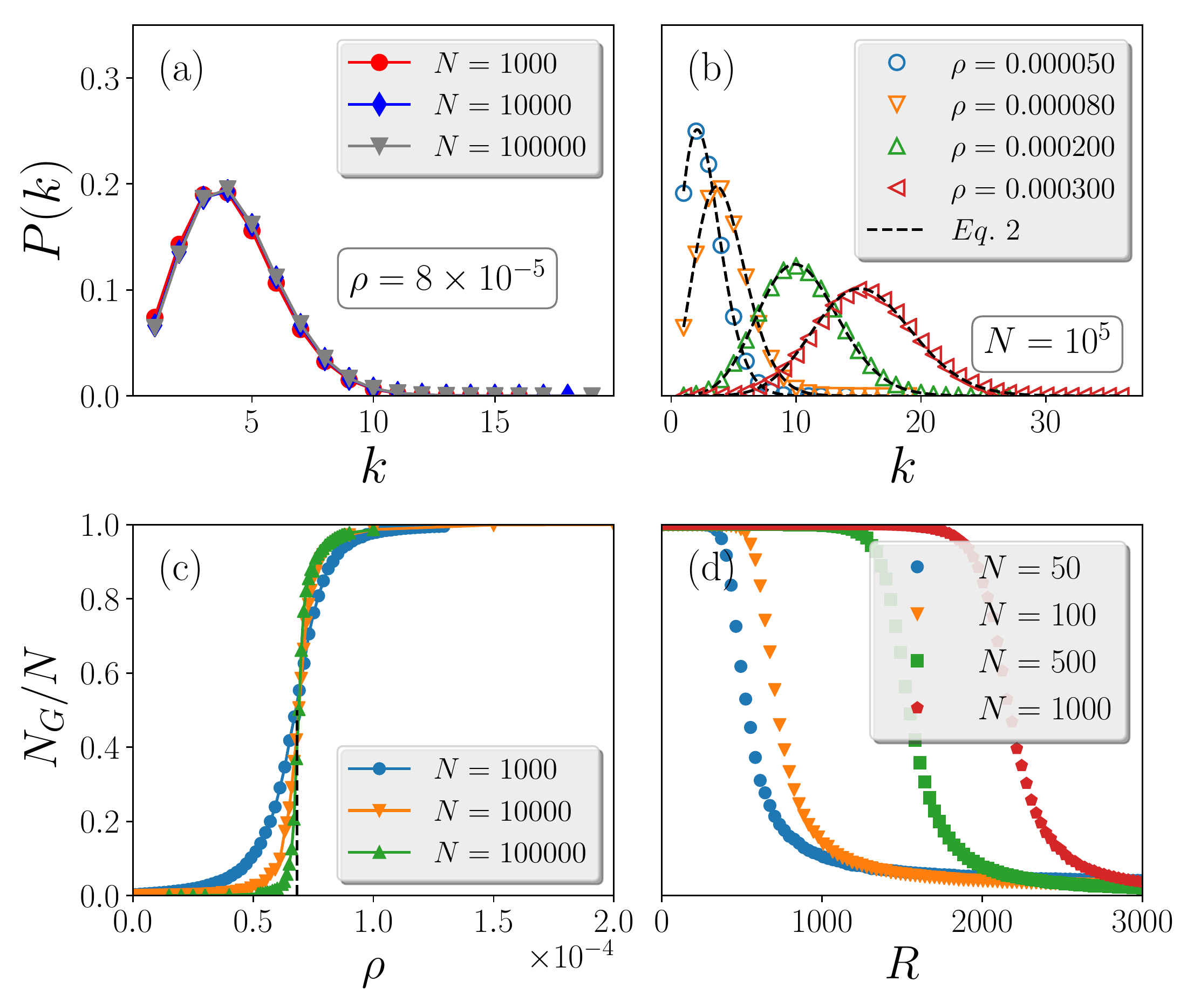}
\end{center}
\caption{\textbf{Degree distribution and emergence of the giant cluster.} (Top panel) $(a)$ The degree distribution $P(k)$ for a fixed density value of $\rho = 8\times 10^{-5}$ and several values of $N$. $(b)$ $P(k)$ for a fixed value of $N$ and several values of $\rho$. (Bottom panel) Relative size of the giant cluster as a function of $(c)$ $\rho$ (density). We see a clear phase transition at $\rho_c\approx6,82\times10^{-5}$ corresponding to $\langle k \rangle_c \approx 3.56$. $(d)$  The appearance of the giant cluster for moderate size networks covering a relatively large area.}
\label{pdk_cmed}
\end{figure}

The first property we analyze is the  degree distribution $P(k)$. As shown in Fig.~\ref{pdk_cmed} (top panels), the degree distribution can be perfectly fitted a Poisson distribution that depends solely on the density of nodes $\rho$:
\begin{eqnarray}\label{eq_pdk_ro}
   P(k) = \frac{e^{-A\rho}(A\rho)^{k}}{k!},
\end{eqnarray}
where $A=5,2\times 10^4$ (see Appendix). This suggests that the quantum internet has a similar structure of a random network. 

Another similarity with random networks model is the existence of a phase transition (see bottom panels of Fig.~\ref{pdk_cmed}) from a disconnected phase to a phase where a giant cluster is present. In the quantum internet model however, this transition is controlled by the density of nodes. We have estimated the critical density to be $\rho_c \approx 6,82\times 10^{-5}$  (see Fig. \ref{pdk_cmed}c), which corresponds to $\langle k \rangle_c \approx 3.5$ (as opposed to $\langle k \rangle_c = 1$ in random networks). Nicely, this critical density is quite small, implying that large areas can be connected by few nodes. For instance, Fig. \ref{pdk_cmed}d shows that the minimum number of nodes necessary to have a connected network in areas comparable to the US or Europe are of the order of $1000$ nodes.

As showed in the Fig.~\ref{pdk_cmed}c, the relative size of the giant cluster displays a second order phase transition with respect to the density $\rho$. In the Appendix we show that $m \equiv \langle N_G \rangle /N$, at the critical density $\rho_c \simeq 6.82\times10^{-5}$, exhibits a power law behavior given by $m \sim (\rho - \rho_c)^{\beta}$, with the associated critical exponent $\beta\simeq 0.2$. Furthermore, we also analyzed the standard deviation of the size of the largest cluster, analogous to the susceptibility, defined by $\chi \equiv \sqrt{\langle N_G^2\rangle - \langle N_G \rangle^2}$, the characteristic cluster size $s^{*}$, the cluster size distribution $n(s)$ and computed the associated critical exponents (see Appendix for more details).

\begin{figure}[t!]
\begin{center}
\includegraphics[width=0.99\columnwidth]{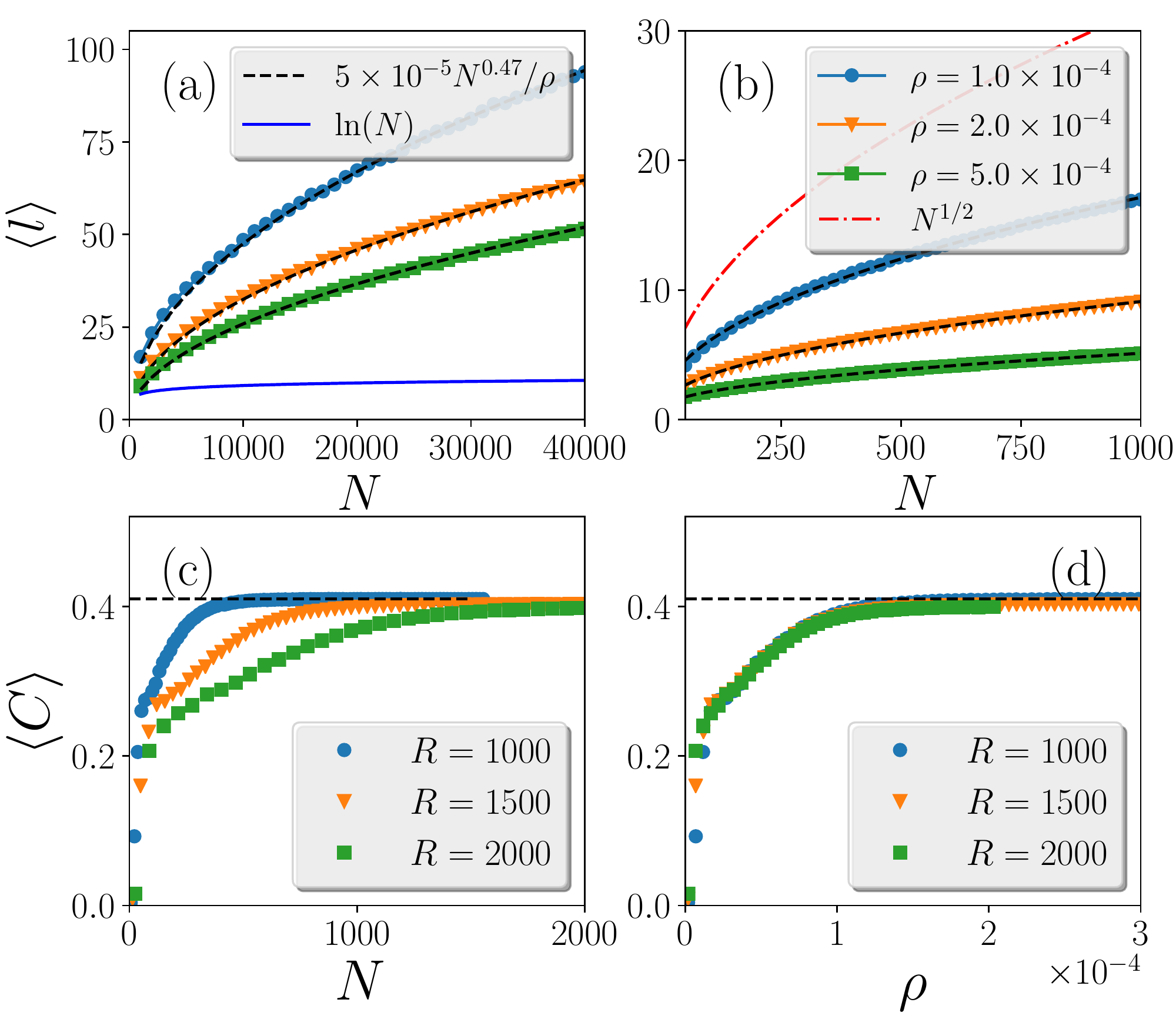}
\end{center}
\caption{\textbf{Average shortest path and average clustering coefficient.} (Top panel) $(a)$ $\langle l \rangle$ as a function of $N$ for various values of $\rho$. $\langle l \rangle$ grows faster than $\ln N$, showing no small-world phenomenon as expected for random networks.
$(d)$ However, the average shortest path for moderate size networks can be relatively small.  (Bottom panel) $(c)$ Average clustering coefficient $\langle C \rangle$ as a function of $N$ for fixed values of $R$. $\langle C \rangle$ grows with $N$ and decreases with $R$. $(d)$ Plotting $\langle C \rangle$ as a function of $\rho$ we see that all curves collapse, showing a universal behavior.}
\label{giant_cluster_earth}
\end{figure}

In spite of the previously discussed similarities between the quantum internet and  random networks, we have observed two important differences. First, as shown in Fig.~\ref{giant_cluster_earth}(a), the photonic quantum network does not display the small world property, since $\langle l \rangle$ grows faster than $\ln N$. We estimated that $\langle l \rangle$ depends of $\rho$ and $N$ following the relation $\langle l \rangle \simeq b N^{\alpha}/\rho$ with $b=5\times 10^{-5}$ and $\alpha \approx 1/2$ as can be seen in the fit of the Fig.~\ref{giant_cluster_earth}(a). Nevertheless, as shown in Fig. \ref{giant_cluster_earth} (b), for moderate network sizes $\langle l \rangle$ is still small.

The average clustering coefficient of the photonic quantum networks also differs from the random network case. As we can see in Fig.~\ref{giant_cluster_earth}(c), $\langle C \rangle$ increases with $N$ independently of the radius $R$ and reach a maximum value $\langle C \rangle \simeq 0.41$. This means that the photonic quantum networks can be classified as very aggregated \cite{barabasi2016network}. Furthermore, as shown in Fig.~\ref{giant_cluster_earth}(d), all curves collapses when we plot $\langle C \rangle$ as function of $\rho$, pointing out the emergence of a universal behaviour. This means that we can describe any curve of the clustering coefficient for any value of $R$ with the same function.

In this article we have proposed a model to study the properties of a quantum internet based on optical fiber technology. Using this model we predicted a phase transition, where there is a critical network density at which a giant cluster suddenly emerges. Crucially, the critical density separating the two phases is quite small, implying that few nodes are needed to hold a fully connected network in realistic areas. We also showed that, even though the generated networks are very aggregated locally, they do not lead to the small-world property. Although this might seem as a negative result, we also showed that for realistic networks sizes, the typical network distances between nodes are small. For instance, in a disk of radius $R=800$km, we would need around $N=1000$ nodes to have a connected network, while keeping the average shortest path length of $\langle l \rangle\approx5$. This has an important implication, for instance, in entanglement distribution. Suppose that the photonic links generating the networks are used to establish entanglement between the nodes (\ie, each link can be seen as an entangled pair of photons). In this case, it would be possible to generate entanglement between any two nodes of the network by performing entanglement swapping on (approximately) five intermediate nodes. 

Our results give a novel perspective to analyze the quantum internet, providing an interdisciplinary bridge between quantum information and network theory. The present contribution should be seen as a starting point towards more complicated models. For instance, here we did not consider that each node can hold a quantum memory that stores the quantum information until the next photonic pulse arrives. Another layer of complexity would be to consider the quantum features of the arriving photons, such as coherence and entanglement. Finally, it would be interesting to consider other technologies such as the use of satellites for quantum communication \cite{Bedington2017,Yin1140,PhysRevLett.120.030501}. 

We thank R. Pereira, C. Argolo, João M. de Araújo and George Moreno for useful discussions about critical exponents. We acknowledge the John Templeton Foundation via the Grant Q-CAUSAL No. 61084, the Serrapilheira Institute (Grant No. Serra-1708-15763), the Brazilian National Council for Scientific and Technological Development (CNPq) via the National Institute for Science and Technology on Quantum Information (INCT-IQ) and Grants No. $423713/2016-7$, No. 307172/2017-1 and No. 406574/2018-9, the Brazilian agencies MCTIC and MEC. DC acknowledges the Ramon y Cajal fellowship, the Spanish MINECO (QIBEQI FIS2016-80773-P, Severo Ochoa SEV-2015-0522), Fundacio Cellex, and the Generalitat de Catalunya (SGR 1381 and CERCA Programme). AC acknowledges UFAL for a paid license for scientific cooperation at UFRN. We thank the High Performance Computing Center (NPAD) and DFTE-UFRN for providing computational resources.
\bibliography{Ref}

\newpage
\section{Appendix}

\section{Average degree, network density and average distance}

\begin{figure}[h!]
\begin{center}
\includegraphics[scale=.33]{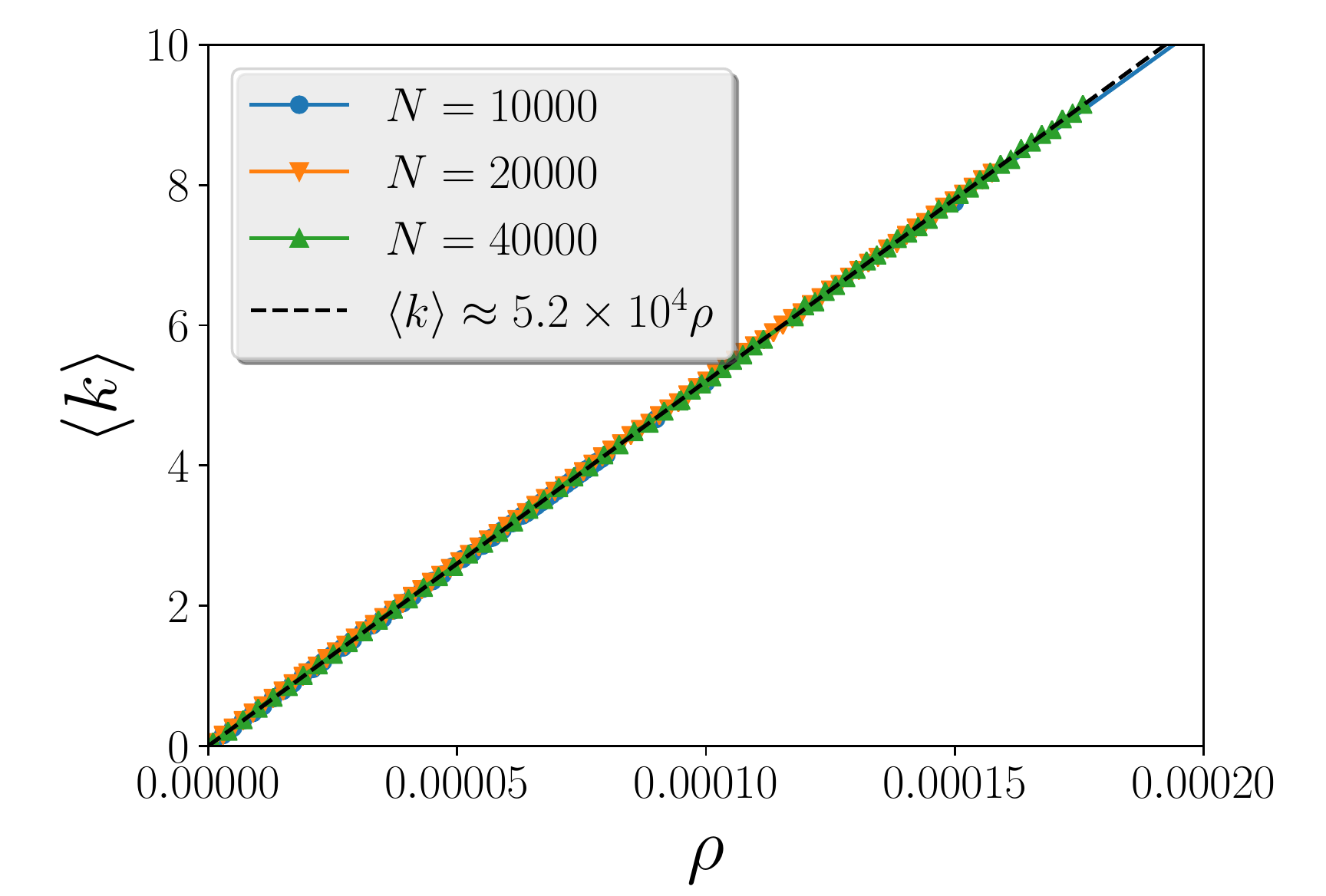}
\end{center}
\caption{\textbf{Linear relation between $\langle k \rangle$ and $\rho$}}
\label{kmed_ro}
\end{figure}

As mentioned in the main text we have numerically derived some relations between $\langle k \rangle$ versus $\rho$ and $\langle l \rangle$ versus $N$. As shown in the Fig.~\ref{kmed_ro}, the average degree is related with the density trough a linear function given by $\langle k \rangle = A\rho$, where $A \simeq 5,2\times 10^{4}$, that is independent of the size $N$ of the network. Because that, all curves of the average connectivity distribution $P(k)$ can be described by the same function as shown in the Fig.~$2$(b) of the main text.

\begin{figure}[h!]
\begin{center}
\includegraphics[scale=.3]{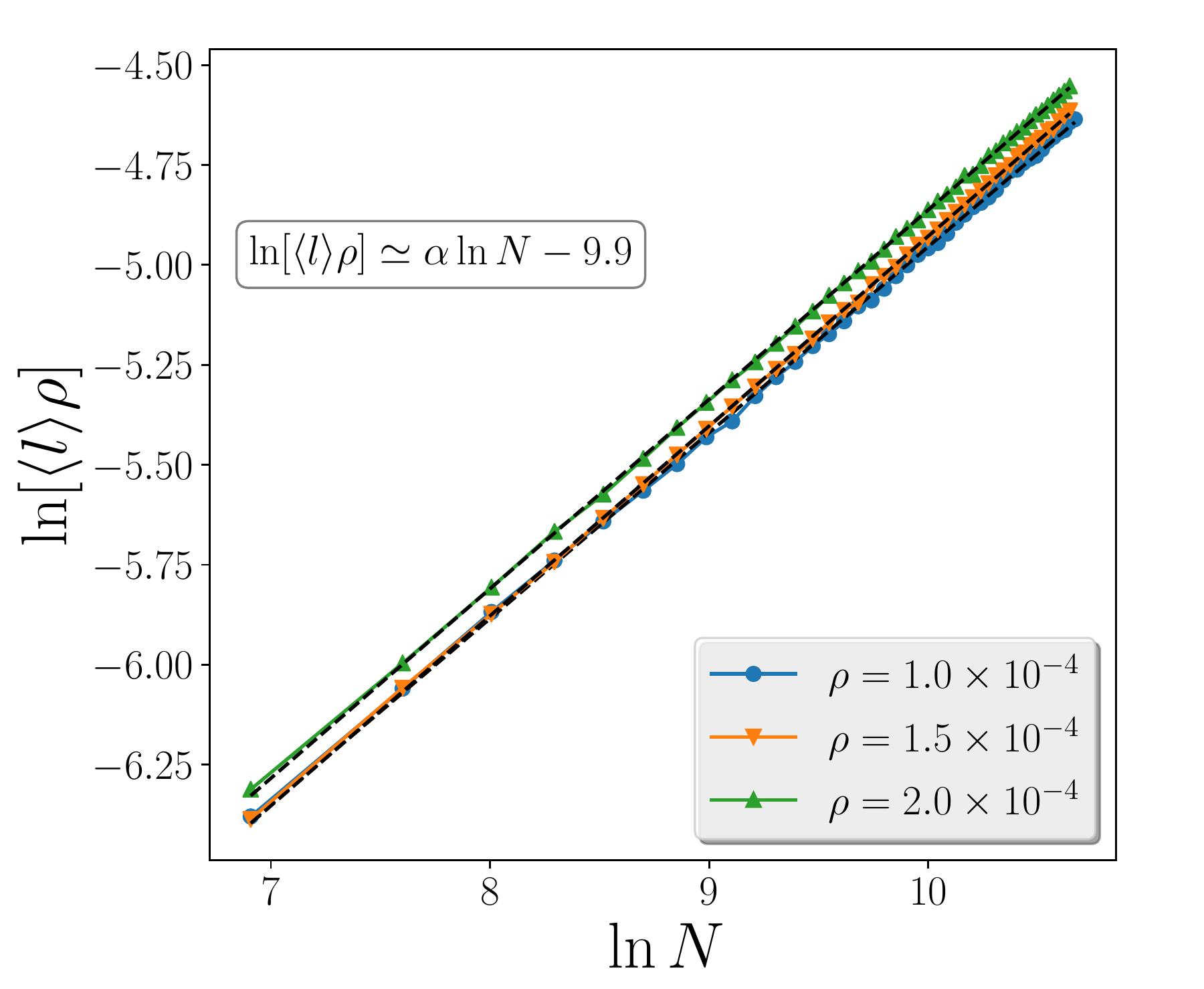}
\end{center}
\caption{\textbf{Linear function between $\langle l \rangle$ and $\sqrt{N}$.} By rescaling the axis $\langle l \rangle \to \ln[ \langle l \rangle \rho]$ and $ N \to \ln N$ we can see that all curves present the same linear behavior given by $\ln[\langle l \rangle \rho] \simeq 0.47 \ln N - 9.9$. From that can easily show that $\langle l \rangle \simeq N^{\alpha} e^{-9.9}/\rho$. This expression can be approximated by $\langle l \rangle \approx 5\times 10^{-5} N^{\alpha}/\rho$ with $\alpha \to 1/2$. This result shows that our model do not generate a small world network.}
\label{mc_linear_N}
\end{figure}

We also have analysed how the average shortest path length $\langle l \rangle$ scales with $N$ for different values of $\rho$. As can be seen in the Fig.~\ref{mc_linear_N}, independent of the density, all curves of $\langle l \rangle$ can be described by a universal function given by $\langle l \rangle \approx 5\times10^{-5} N^{\alpha}$ with $\alpha  \approx 0.5$. This result show our model do not displays a small world phenomenon.

\section{Phase transition at different values of the physical parameters}

\subsection{Changing the optical fiber loss}

\begin{figure}[h!]
\begin{center}
\includegraphics[scale=.33]{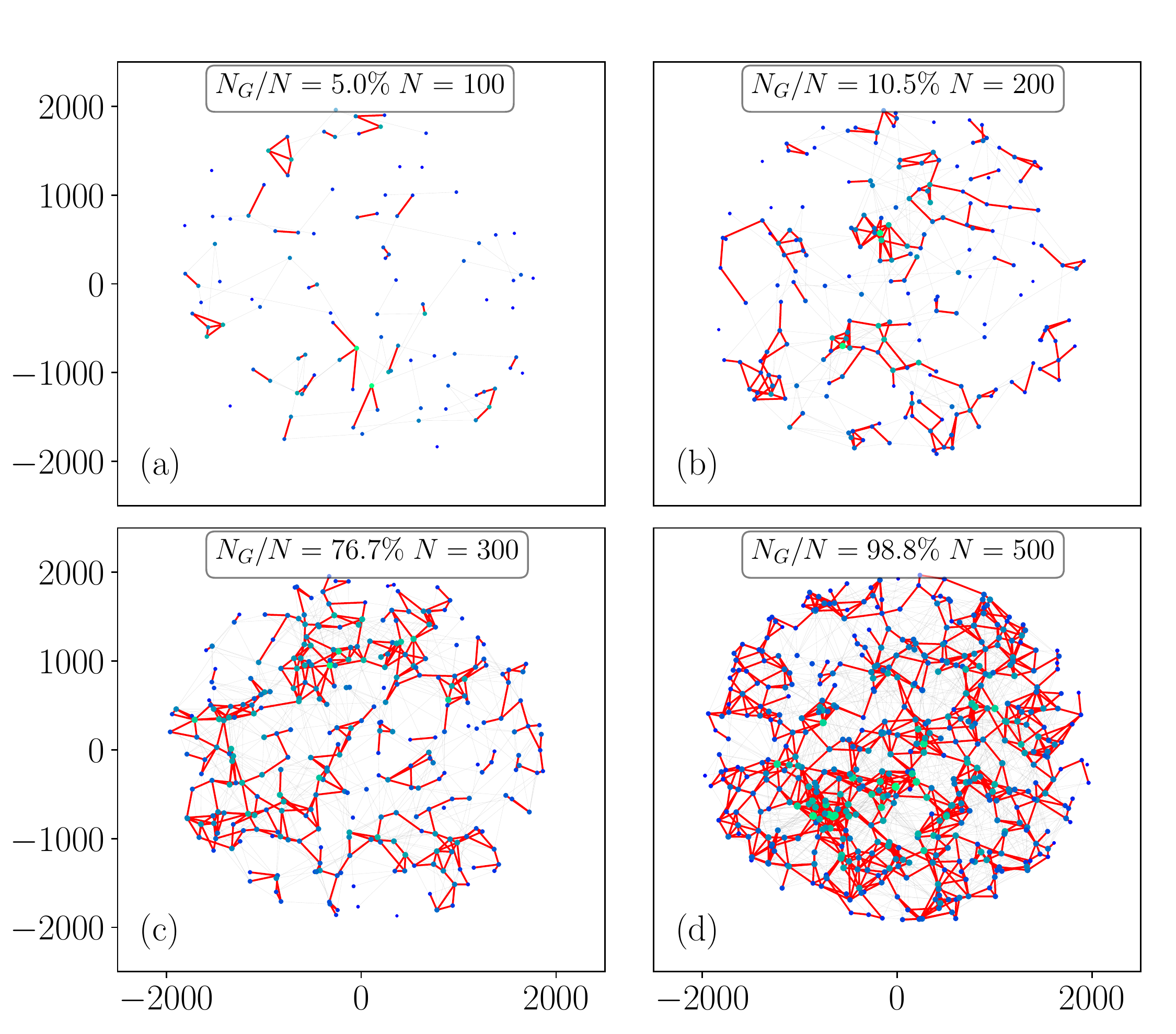}
\end{center}
\caption{\textbf{Samples from the quantum internet} with $\gamma = 0.095$ dB/km. The grey edges represents the fiber-optics networks generated at step 1 (see main text). The red edges show the photonic links established in step 2. Greener (bluer) nodes are more (less) connected, following a Poisson distribution. $N_G$ refers to the number of nodes belonging to the biggest cluster in the network, and $N$ the total number of nodes. The plots considered $R=2000$ km and show that the biggest cluster consists of $98,8\%$ of the nodes when $N=500$.}
\label{net_gamma2}
\end{figure}

\begin{figure}[h!]
\begin{center}
\includegraphics[scale=.33]{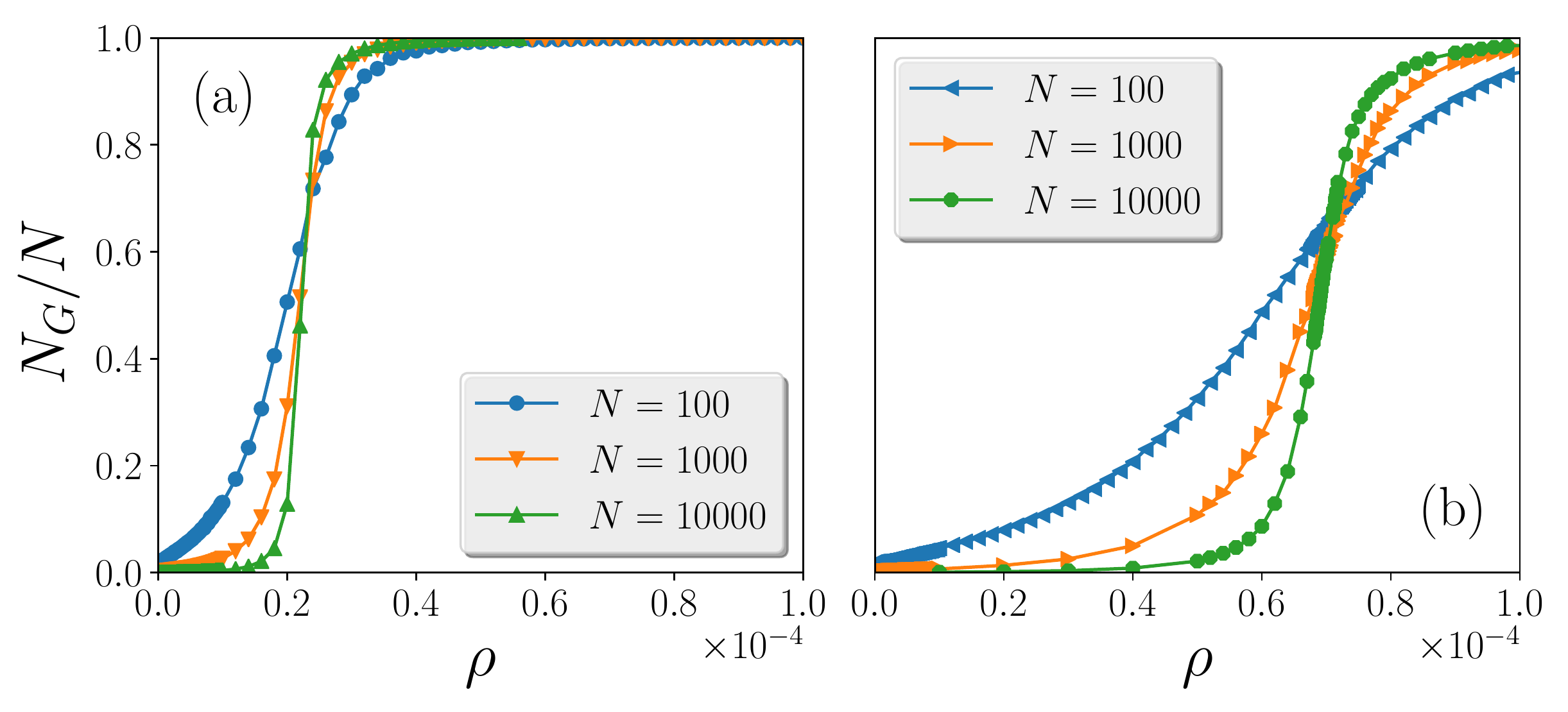}
\end{center}
\caption{\textbf{Appearance of the giant cluster for different values of $\gamma$}. Network generated for $n_p = 1000$, $\alpha L = 226$ km, $\beta =1$, $10^3$ realizations,  $(a)$ $\gamma = 0.095$ dB/km and $(b)$ $\gamma = 0.2$ dB/km. As show above the appearance of the giant cluster associated still present a phase transition, but for different value of $\rho_c$.}
\label{NG_gamma2}
\end{figure}

To illustrate the fact that the phase transition happens irrespective of the optical fiber being employed, below we show additional results by considering a better optical fiber with $\gamma = 0.095$ dB/km. By using the same parameters  $(n_p=1000, \alpha L = 226\text{ km}, \beta = 1)$ we still observe a phase transition in the relative size of the giant cluster \ref{NG_gamma2}, but now in a different critical density $\rho_c$. As expected, the phase transition happens at a smaller critical density. The results are summarized in Figs. \ref{net_gamma2} and \ref{NG_gamma2}. 

\subsection{Changing the value of $n_p$}

\begin{figure}[h!]
\begin{center}
\includegraphics[scale=.33]{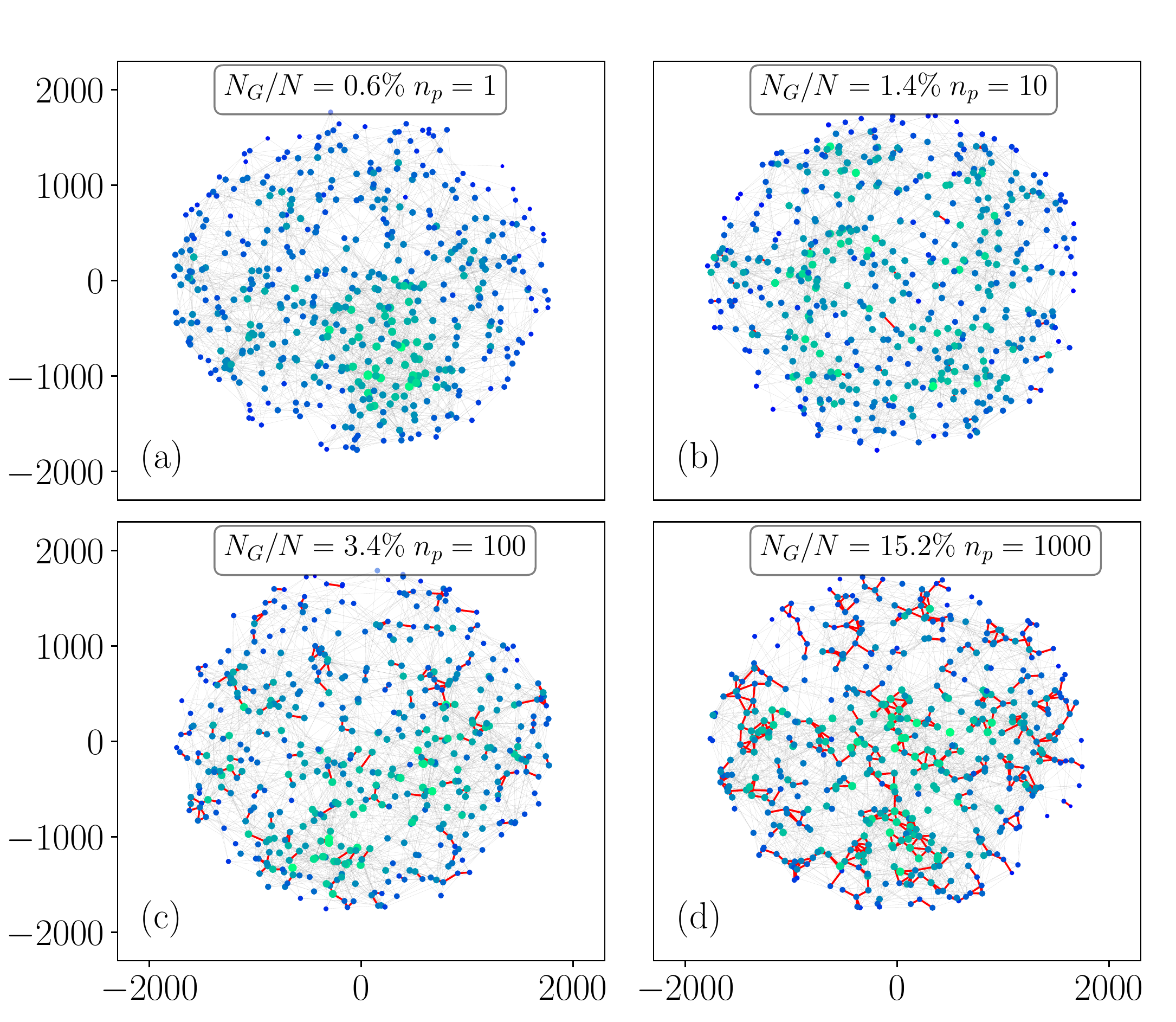}
\includegraphics[scale=.33]{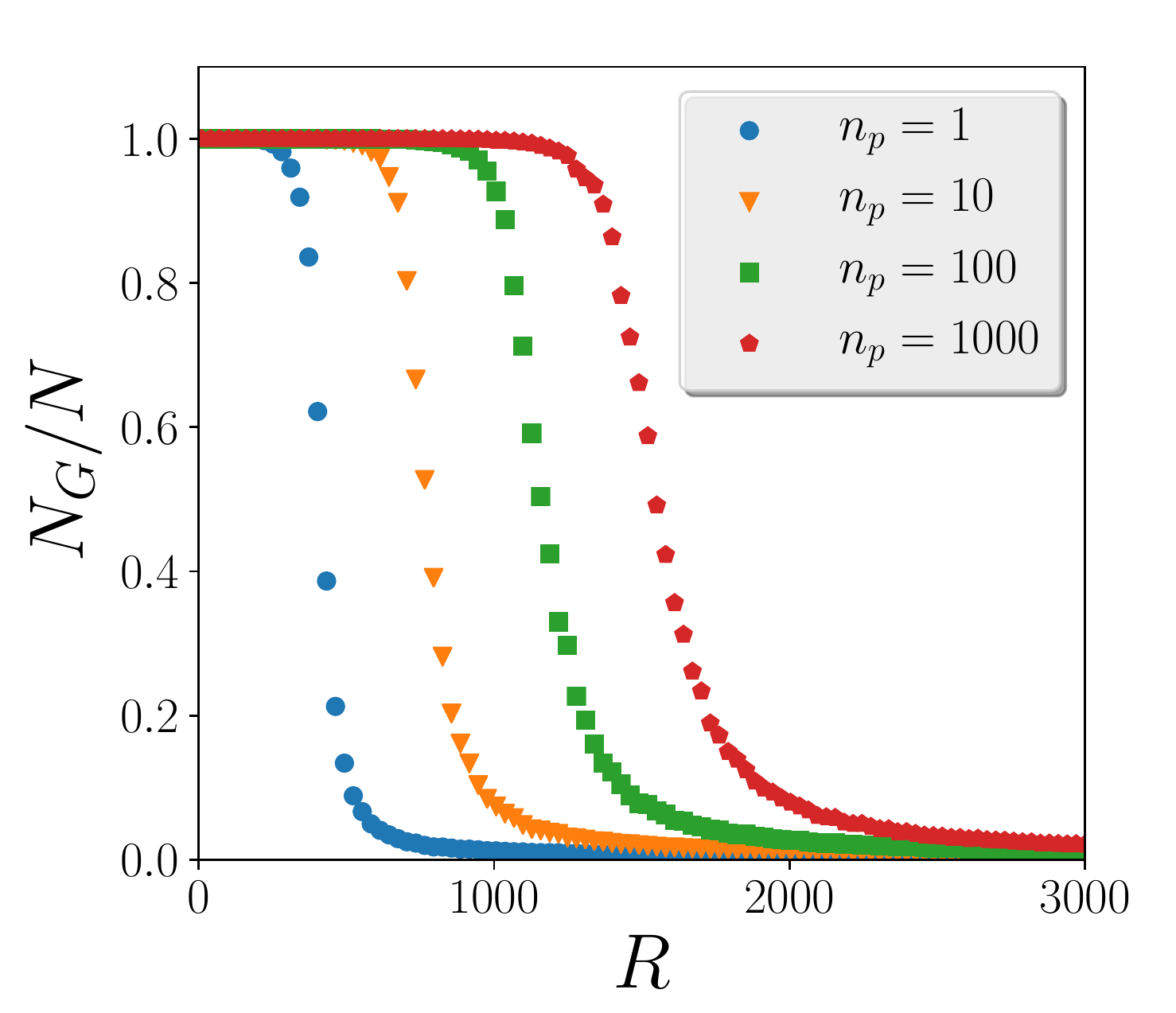}
\end{center}
\caption{(Top) \textbf{Samples from the quantum internet} with $\gamma = 0.2$ db/km, $R = 1800$ km, $N=500$ and different values of $n_p$. The grey edges represents the fiber-optics networks generated at step 1 (see main text). The red edges show the photonic links established in step 2. Greener (bluer) nodes are more (less) connected. $N_G$ refers to the number of nodes belonging to the biggest cluster in the network, and $N$ the total number of nodes. (Botton) \textbf{Appearance of the giant cluster as function of $R$ for $N=1000$ and several values of $n_p$.}}
\label{net_np}
\end{figure}

Another important parameter of the our model is $n_p$. Once more, to illustrate the fact that the phase transition happens independently of the precise values, we consider different values of $n_p$. Clearly, by changing this value we also change the critical point $\rho_c$ of the phase transition. The results are shown in Figs.~\ref{net_np}.

\section{Phase Transition and Critical exponents of the quantum network}

This section presents the numerical estimation of the critical exponents related with the quantum network model proposed in the main text. Our analysis is limited by network sizes of with up to $10^5$ nodes due to computational cost. However, it suffices for comparing the statistical properties of other graph-based models with the proposed one, as shown in what follows. 

\subsection{Critical points and phase transitions}

Physical systems may present abrupt changes in macroscopic behaviour, termed as phase transitions \cite{gene}, such as the emergence of a giant cluster connecting most of the sites as discussed in the main text. As shown, this clustering transition takes place only above a certain density of nodes known as the critical point, in our case $\rho_c$. Below this density there is no dominating cluster appearance with size of the order of the whole network. Qualitatively, those universal behaviours can be well expressed in terms of power laws of the form $\langle u_i(t) \rangle \sim (t-t_c)^{\alpha_i}$ for the average of some macroscopic quantities $u_i$ in the near vicinity of some critical point $t \approx t_c$.

The order parameter $m$ is one of the most important quantities in the analysis of a phase transition. We use $m = \langle N_G\rangle /N$, which is the relative size of the largest cluster. When the order parameter faces a sudden jump at the critical point the transition is said to be first order or discontinuous. On the other hand, if the order parameter varies continuously near the critical point the transition is second order. Identifying the nature of changing properties plays a crucial role in phase transition phenomena. It is generally done by determining the universality class of the transitions. The values of the critical exponents define the universality class of the transition under investigation. However, such analysis is computationally demanding as a large amount of individual realizations of the network is required in order to obtain good statistical ensemble to calculate accurately the critical exponents.   

Broadly speaking, critical exponents depict the behavior of physical quantities in regions close to continuous phase transitions driven by a control parameter. By continuity we mean that the growth of the largest cluster is continuous, which is proven to be the case in models based on random graphs \cite{riordan,grass}. It is an important trait, once discontinuous phase transitions can not be characterized by means of critical exponents. Although not formally proved, it is believed that the critical exponents are universal, meaning that they do not depend on microscopic details of the physical system, but only on some of its general features. 

\begin{figure}[t!]
\begin{center}
\includegraphics[scale=.33]{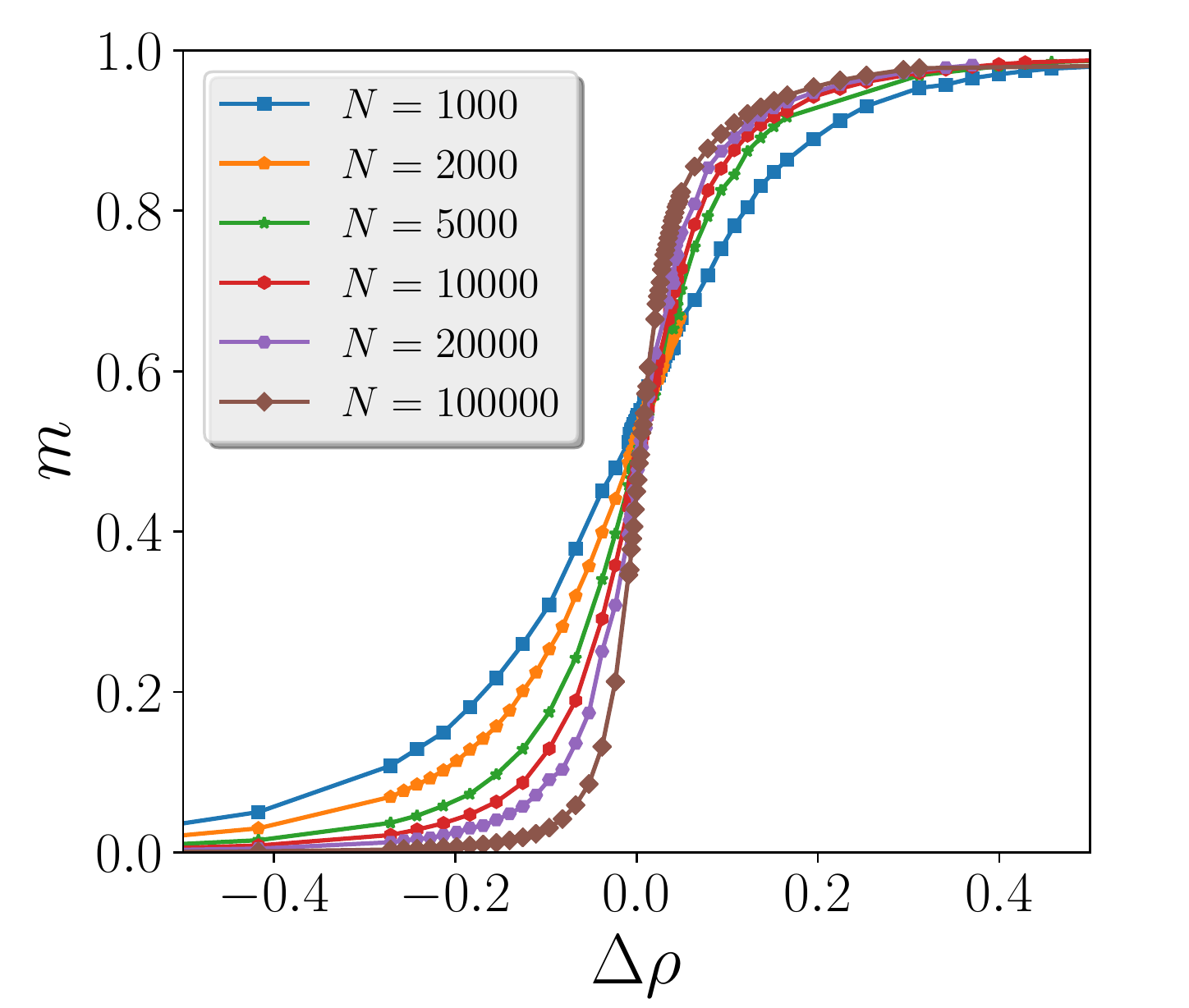}
\end{center}
\caption{\textbf{Emergence of the giant cluster.} Relative size of the giant cluster as a function of  $\rho$ (density) for diverse values of number of nodes $N$. The crossing of the curves indicates finite size effects.}
\label{cross}
\end{figure}

As already discussed, the variable that controls the phase transition in our study is the density of nodes $\rho$. In order to numerically estimate the critical exponents with good accuracy one needs first to compute the density threshold $\rho_c$ of our system. This is primarily done using the interception method in which the value of $\rho_c$ is taken to be the one corresponding to the crossing point for the plot of $m$ as a function of $\rho$ for diverse values of $N$; the interchanging in the behaviour of small and large system size happening due to finite size effects. In our simulations we use the minimum of $1000$ statistical samples for measuring $m$ for various network size $N$ and we can make a first crude estimation that $\rho_c \approx 0.000068(2)$. 

\begin{figure}[t!]
\begin{center}
\includegraphics[scale=.35]{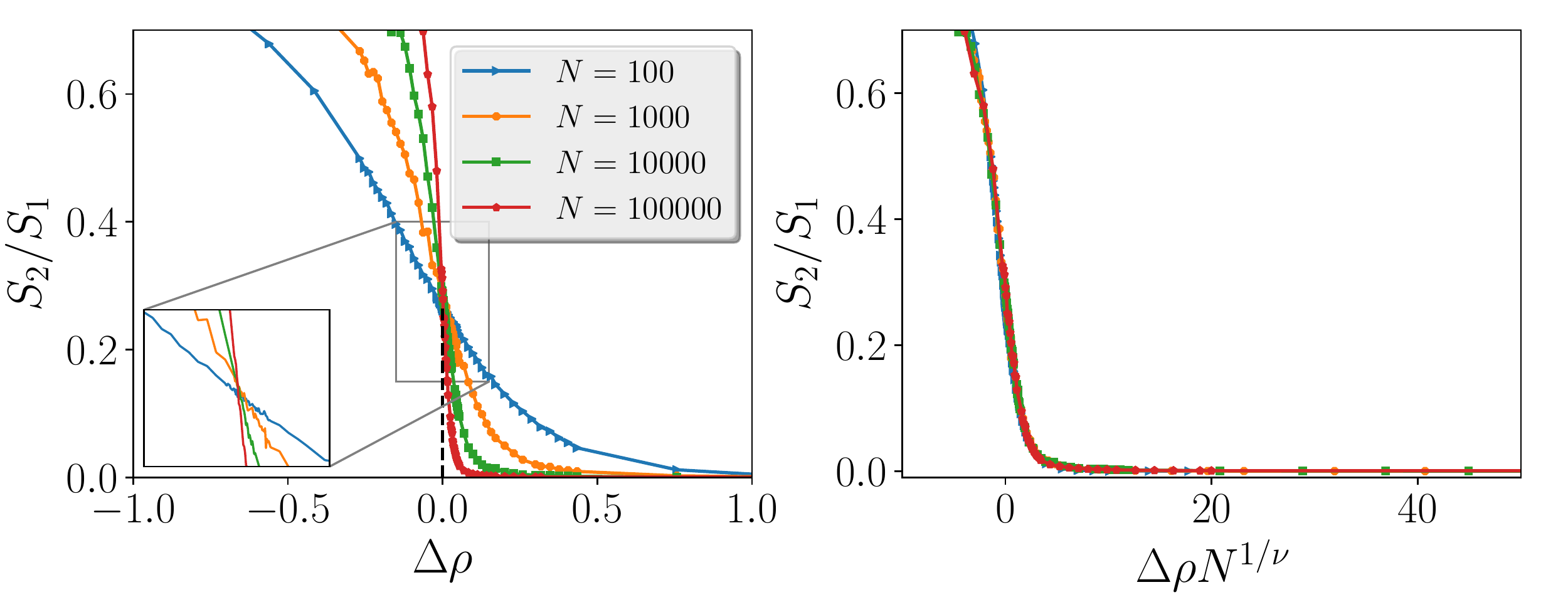}
\end{center}
\caption{ \textbf{Critical density estimation and the critical exponent $\nu$.} (a) $S_2/S_1$ in function of $\Delta\rho$ for diverse values of $N$, where $S_1$ and $S_2$  are the sizes of the largest and the second largest cluster, respectively. The critical density is estimated at the crossing point $\rho_c = 0.0000682$. (b) The critical exponent $\nu$ is estimated by means of the quality of the data collapse of the curves $S_2/S_1$ vs $\Delta \rho N^{1/\nu}$ for diverse values of the system size $N$ with $\nu = 2.78$.}
\label{binder_nu}
\end{figure}

However, a more precise estimation of $\rho_c$ is generally done by using the Binder parameter or Binder cumulant defined as $\text{U}_N = 1 - \frac{\langle m^4 \rangle_N}{\langle m^2 \rangle_N^2}$, being the kurtosis of the order parameter $m$. Notwithstanding, smoother results for the estimation of the critical point have been achieved for the ratio $S_2/S_1$, where $S_2$ and $S_1$ are the size of the second largest and the largest cluster, respectively. In fact, in Ref. \cite{liu} it was shown that the ratio $S_2/S_1$ is similar to the Binder cumulant, which is known to be universal at the critical point. Therefore, we can estimate the critical point plotting $S_2/S_1$ vs $\rho$ and identifying the crossing point at $\rho_c = 0.0000682$, as shown in Fig. \ref{binder_nu} (a). This means that at the critical point the ratio $(S_2/S_1)|_{\rho = \rho_c}$ is independent of the system size $N$, hence the crossing at $\rho = \rho_c$. Furthermore, using the practical finite size assumption of how the characteristic length scales around the critical point $N \propto (\rho - \rho_c)^{-\nu}$ (to be discussed ahead), we can collapse the curves $S_2/S_1$ for different sizes $N$ plotting $S_2/S_1$ vs $\Delta \rho N^{1/\nu}$ using the right values of the critical exponent $\nu$, where $\Delta \rho = (\rho - \rho_c)/\rho_c$. In Fig. \ref{binder_nu} (b) we show the data collapse with $\nu = 2.78$. The quality of the match, apart from some finite size effects, is an heuristic way to determine the critical exponent $\nu$.  

\subsection{Finite-size scaling}

When dealing with any macroscopic system with a vast number of degrees of freedom, we invariably need to make approximations/simulations of a smaller model system due to the limited amount of computational resource. This introduces systematic errors called finite size effects, e. g., the crossing of the curves in the phase diagrams in function of $N$ shown in Fig. \ref{cross}. It means that a system with $N$ parties undergoes a true phase transition only when $N \to \infty$. A deeper understanding of the phenomena, therefore, requires the extrapolation to an infinite system, usually done by means of a number of simulations at different system sizes. Withal, within the field of statistical mechanics there exists an elegant and useful theoretical paradigm to perform this extrapolation for phase transitions, such as the finite size scaling.  

The scaling theory seeks to provide a simplified description of a system near a critical point, where many properties depend strongly on the system size $N$. This $N$-dependence allows one to determine the critical point and estimate exponents using data for different system sizes. It relies on the assumption that when we approach the critical point, there exists a characteristic length scale following $\xi \propto (\rho - \rho_c)^{-\nu}$. Therefore, the size-dependence is related to the ratio $N/\xi$, conventionally represented by the term $(\rho - \rho_c)N^{1/\nu}$ \cite{fisher}. 

In order to characterize the phase transition of our model, we compute the following critical exponents:
\begin{eqnarray}
 m  &\sim& \Delta\rho^\beta,\\
 \chi &\sim& \Delta\rho^{-\gamma\prime},\\
  N &\sim& \xi \sim \Delta\rho^{-\nu}, \\ 
 s^{*} &\sim& \Delta\rho^{-1/\sigma},\\
n(s) &\sim& s^{-\tau}|_{\Delta\rho = 0},
\label{ce}
\end{eqnarray}
where $\chi$ is the standard deviation of the size of the largest cluster, similar to the susceptibility definition, $s^*$ is the characteristic cluster size, $n(s)$ is cluster size distribution. We use the powerful finite size scaling ansatz to compute the corresponding critical exponents.

\begin{figure}[t!]
\begin{center}
\includegraphics[scale=.40]{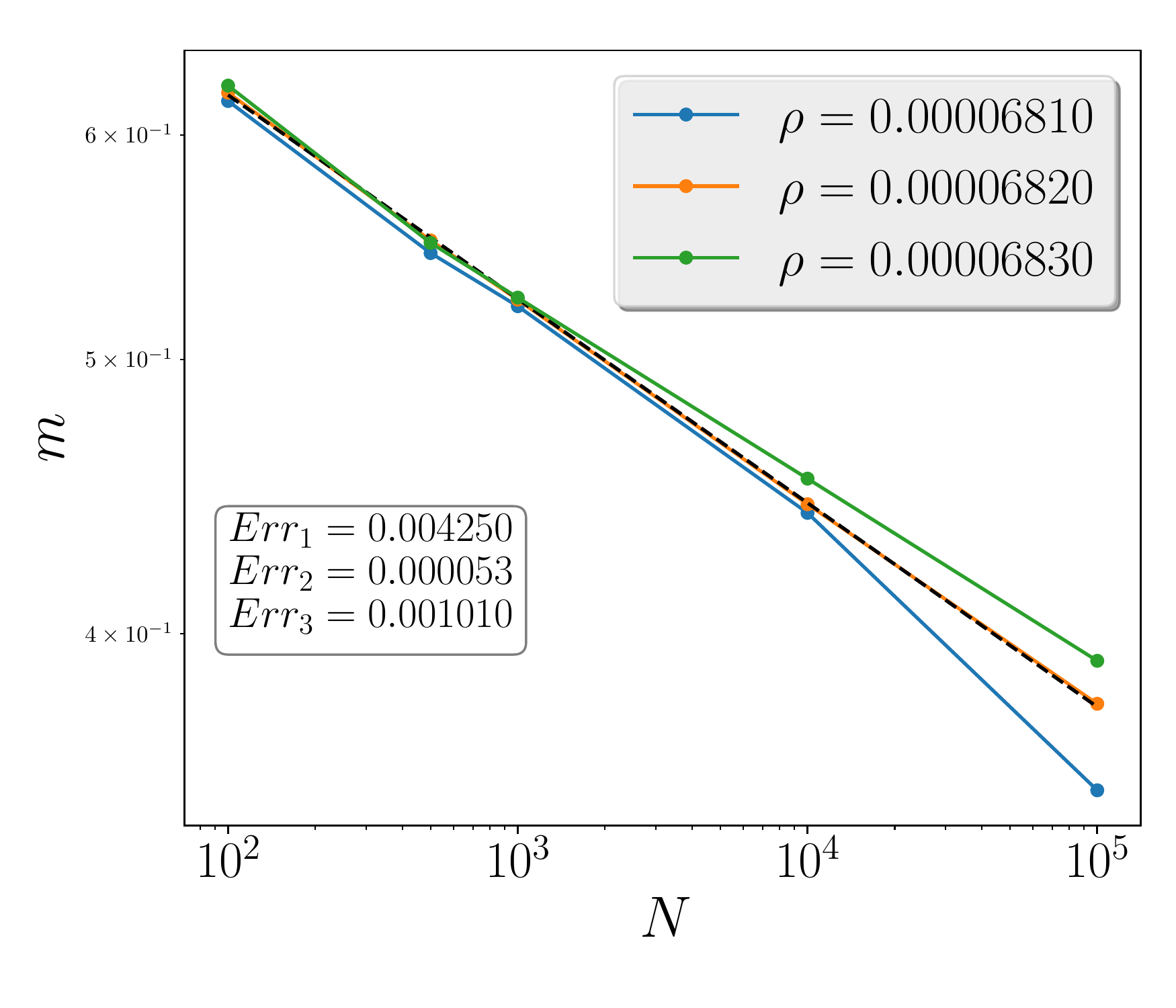}
\end{center}
\caption{\textbf{Critical density estimation and the critical exponent ratio $\beta/\nu$.} Estimation of the critical density by means of the best power law behavior for the FSS ansatz $m \propto N^{- \beta/\nu}f_{\beta} [(\rho - \rho_c)N^{1/\nu}]$. For $\rho = \rho_c$ we get the power law behavior $m \propto N^{- \beta/\nu}f_{\beta} (0)$. The black dashed line shows the best fit, retrieving $\beta /\nu = 0.071$.}
\label{beta_ni_all}
\end{figure}

Clearly, we face the numerical limitation of working only with finite size system. As can be seen in Fig. \ref{cross}, plotting $m$ as a function of $\rho$ for different system size indicates that, before the transition point, $m$ remains smaller for higher values of $\rho$ with increasing $N$. This is typical of the order parameter following the finite size scaling (FSS) ansatz discussed above, expressed as
\begin{eqnarray}
m(\rho,N) &\sim& N^{-\beta/\nu} f_{\beta}(\Delta \rho N^{1/\nu}]),
\label{fss}
\end{eqnarray}
where $f_{\beta}(z)$ is analytic at all finite values of $z$, meaning that the only singularity might come from the critical point. It is also said to be the universal scaling function of $m$. The negative sign in the exponent $N^{-\beta/\nu}$ accounts for the observation that $m$ remains smaller for larger values of $\rho$ for increasing $N$, as discussed above.

\begin{figure}[t]
\begin{center}
\includegraphics[scale=.35]{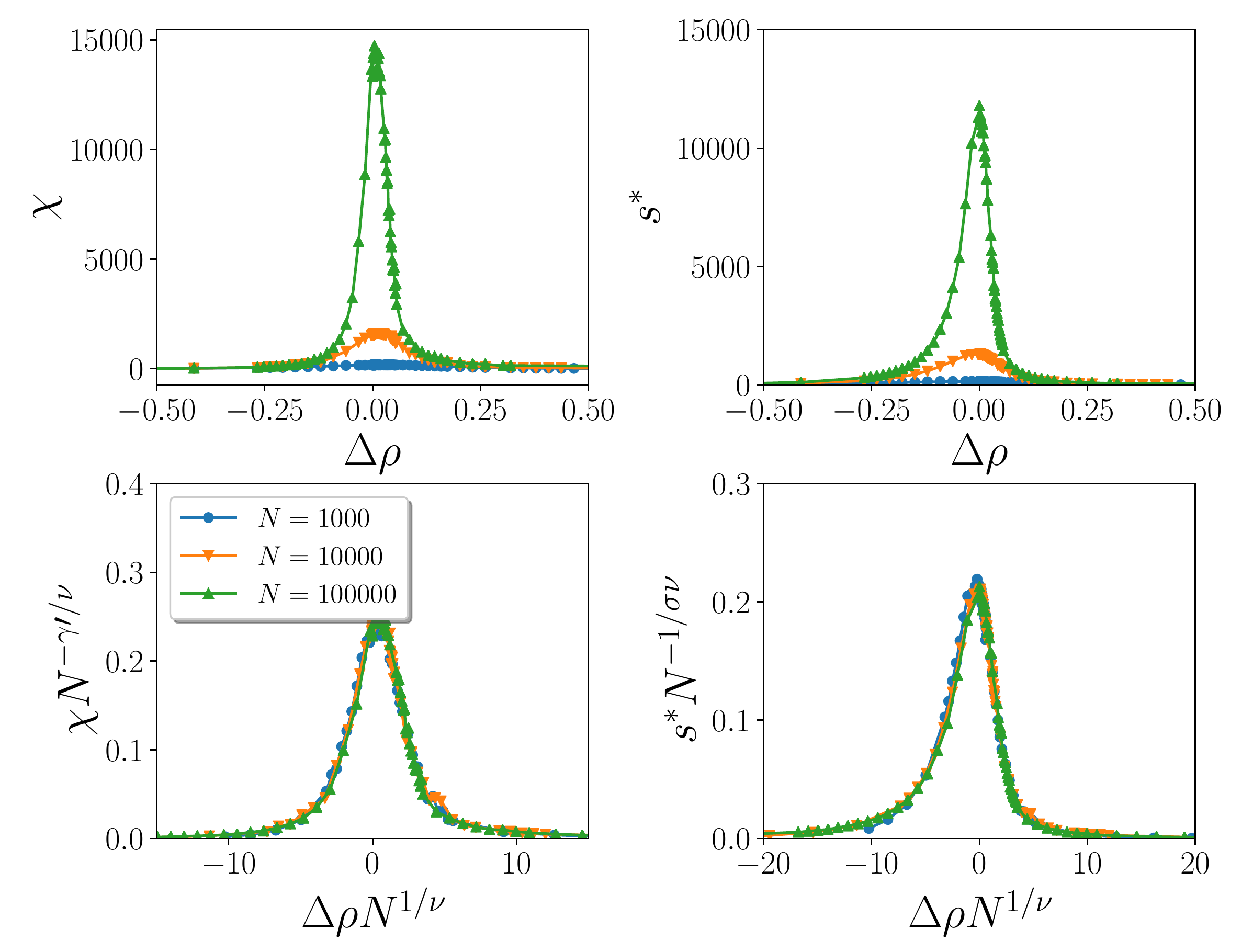}
\end{center}
\caption{\textbf{Data-collapse check.} (a) Plot of $\chi$ in function of $\Delta \rho$ and (b) the corresponding data-collapse into a master curve for the plot $\chi N^{-\gamma\prime/\nu}$ versus $(\rho - \rho_c)N^{1/\nu}$. (c) Plot of $s^{*}$ in function of $\Delta \rho$ and (d) the corresponding data-collapse into a master curve for the plot $s^{*} N^{-1/\sigma \nu}$ versus $(\rho - \rho_c)N^{1/\nu}$. The critical exponents and exponent ratios are: $\nu = 2.78$, $\gamma\prime/\nu = 0.96$, $1/\sigma \nu = 0.94$}\label{colapso}
\end{figure}

Using $\rho= \rho_c$ in Eq. \ref{fss} yields 
\begin{eqnarray}
m(\rho,N) &\sim& N^{-\beta/\nu} f_{\beta}(0),
\label{fss2}
\end{eqnarray}
where $f_{\beta}(0)$ is just a constant. Therefore, the order parameter decays as a power law of $N$, meaning that the log-log plot of $m$ in function of $N$ is a straight line and the slope is the critical exponent ratio $\beta / \nu$. It is also a way of estimating the critical density, guessed with the crossing of the ratio $S_2/S_1$, as shown in Figs.~\ref{binder_nu}(a). The curve for $\rho = 0.0000682$ in Fig.~\ref{beta_ni_all} is the one showing best agreement with a power law output, therefore $\rho_c = 0.0000682$, and using the linear relation between $k$ and $\rho$ shown in the main text, it means $\langle k \rangle_c = 3.562$. The corresponding critical exponent ratio $\beta/\nu = 0.071$ is just the slope of the log-log plot of the line fitting the curve for $m \propto N$ for $\rho_c$. The best slope was chosen to be the one with the lowest $L_2$-like error, see Fig. \ref{beta_ni_all}. 

\begin{figure}[ht]
\begin{center}
\includegraphics[scale=.4]{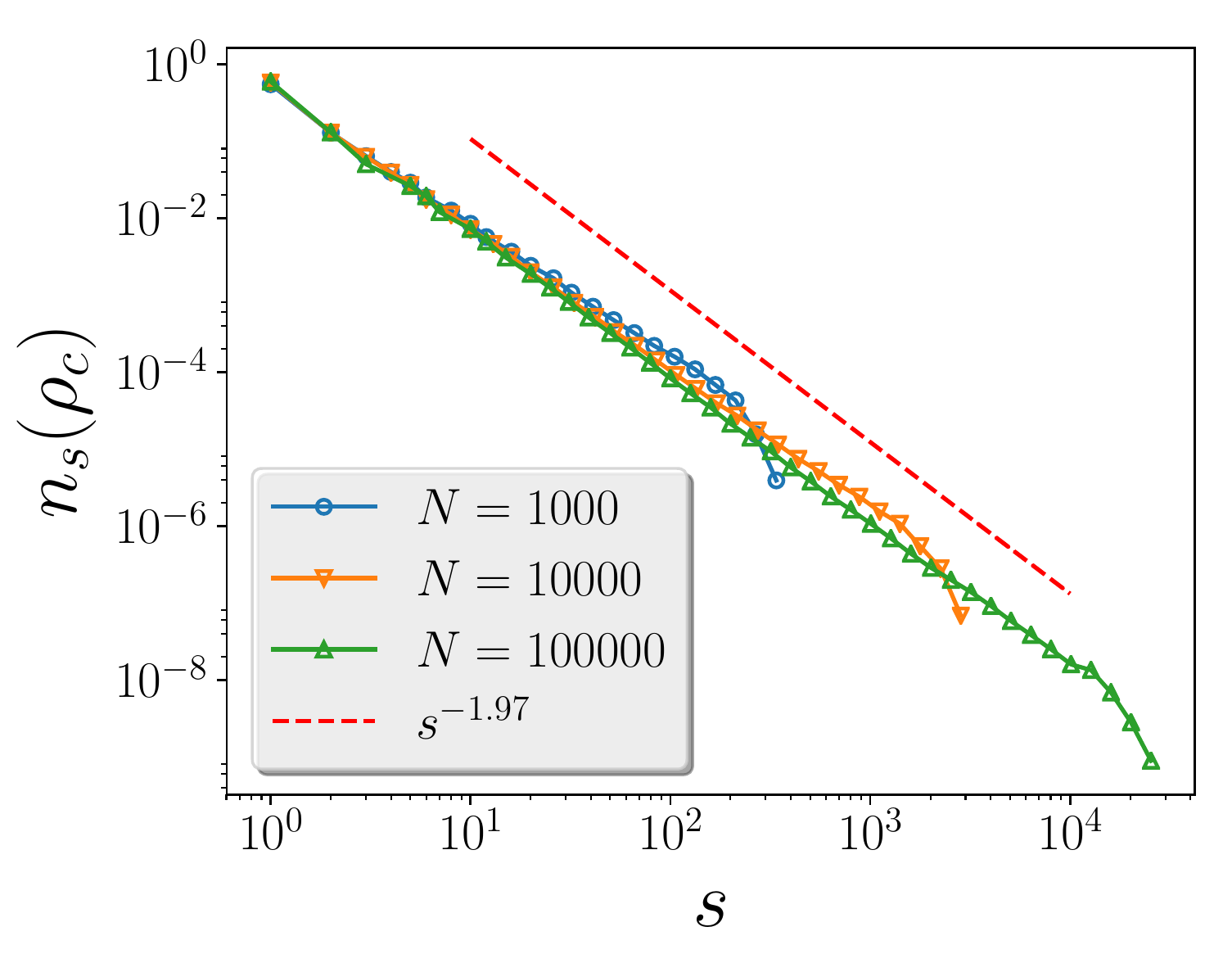}
\end{center}
\caption{\textbf{Critical exponent $\tau$.} Plot of $n(s)$ for distinct values of $N$ in the log-log scale to review $\tau= 1.97$. The red dashed line is a guide for the eye.}
\label{tau}
\end{figure}

Another important critical exponent is related with the susceptibility $\chi(\rho,N)$, which can be defined as the ratio of the change in the order parameter $\Delta m$ and the change in the density $\Delta \rho$ for the corresponding change $\Delta m$. In the limit $\Delta \rho \to 0$ it becomes the derivative of the order parameter $m$ with respect to $\rho$ in a similar way as the paramagnetic-ferromagnetic transition. Here, we compute the standard deviation of the size of the largest cluster $N_G$, $\chi = \sqrt{\langle N_G^2\rangle - \langle N_G \rangle^2}$ \cite{bastas,cho}. Again we face the numerical limitation solved only for $N \to \infty$. To cope with that, we assume once more the following finite size scaling 

\begin{eqnarray}
\chi(\rho,N) &\sim& N^{\gamma\prime/\nu} f_{\chi}(\Delta \rho N^{1/\nu}),
\label{fss3}
\end{eqnarray}
where $f_{\chi}(z)$ is the universal scaling function of $\chi$. For a finite system, we obtain rounded peaks rather than true divergences ($N \to \infty$). The peaks narrow and increase in height as $N$ is increased as shown in Fig. \ref{colapso} (a), hence the positive power $\gamma\prime/\nu$ in Eq. \ref{fss2}. For $\rho = \rho_c$, this increasing in the susceptibility obeys a power-law $\chi_{\text{max}} \propto N^{\gamma\prime/\nu}$. In this way, the log-log plot of $\chi$ versus $N$ at $\rho = \rho_c$ must be a power-law and the slope is the critical exponent ratio $\gamma\prime/\nu = 0.96$. Again, the validity of Eq. \ref{fss3} implies that when we plot $\chi N^{-\gamma\prime/\nu}$ versus $(\rho - \rho_c)N^{1/\nu}$ all the distinct curves in Fig. \ref{colapso} (a) will collapse into an universal curve for the right values of the critical exponents $\gamma\prime$ and $\nu$, as shown in Fig. \ref{colapso} (b).

In a similar fashion, we define the finite size scaling hypothesis for $s^*$

\begin{eqnarray}
s^{*}(\rho,N) &\sim& N^{1/\sigma\nu} f_{\sigma}(\Delta \rho N^{1/\nu}),
\label{fss4}
\end{eqnarray}
where $f_{\sigma}(z)$ is the universal scaling function of $\sigma$. Analogously to the susceptibility, the average cluster size $s^*$ presents a peak at the critical density $\rho_c$, becoming narrower and higher as the system size $N$ is increased, indicating once more a divergence when $N \to \infty$, see Fig. \ref{colapso} (c). The log-log plot of $s^*$ at $\rho = \rho_c$ is once more used to determine the corresponding critical exponent quantity $1/\sigma \nu = 0.94$. Exactly at the critical point, all these curves must collapse in a master curve when one plots $s^{*} N^{-1/\sigma \nu}$ versus $(\rho - \rho_c)N^{1/\nu}$ with the right critical exponent values for $\sigma$ and $\nu$, as shown in Fig. \ref{colapso} (d).

Although a finite size scaling is possible for the cluster size distribution $n(s)$ \cite{reka}, we made use of a direct log-log plot of $n(s)$ versus $s$ as in Eq. \ref{ce} to extract the critical exponent $\tau$, which is the corresponding slope, see Fig. \ref{tau}. In this manner, we obtain $\tau = 1.97$.

Another important remark is that we could not perform a good data-collapse for the order parameter curves shown in Fig. \ref{cross}, probably due to the strong dependence of $\beta/\nu$ with the estimation of the critical density $\rho_c$. Therefore, it paves the way for a dedicated work using larger network size $N$ and deeper investigation of $\rho_c$ to determine the universality class of this phase transition, with seems to fall, based on the considerably abrupt behaviour of the phase diagram in Fig. \ref{cross} for large $N$, somewhere between the usual and explosive percolation classes \cite{lee2,dsouza}.

\end{document}